\documentclass[12pt]{article}

\pdfoutput=1

\usepackage[bulletsep]{collref}
\usepackage{amssymb,graphicx}
\usepackage[intlimits]{amsmath}
\usepackage{bbm}
\usepackage[small]{subfigure}
\usepackage{wick}

\usepackage{MnSymbol}

\usepackage{hyperref}

\ifx\hypersetup\sadfkjashdfkxja\else
\hypersetup{pdftitle={}}
\hypersetup{pdfauthor={}}
\hypersetup{pdfsubject={}}
\hypersetup{pdfkeywords={}}
\hypersetup{plainpages=false}
\hypersetup{bookmarksnumbered=true}
\hypersetup{pdfstartview=FitH}
\hypersetup{pdfpagemode=UseNone}
\hypersetup{colorlinks=false}
\hypersetup{citebordercolor={.5 1 .5}}
\hypersetup{urlbordercolor={.5 1 1}}
\hypersetup{linkbordercolor={1 .7 .7}}
\fi

\def\be{\begin{equation}}
\def\ee{\end{equation}}
\def\bsp{\be\begin{split}}

\def\a{\alpha}

\def\d{\delta}
\def\e{\epsilon}
\def\m{\mu}

\def\r{\rho}
\def\l{\lambda}
\def\t{\tau}

\def\T{\theta}

\def\p{\partial}

\makeatletter

\newcommand{\Rmnum}[1]{\expandafter\@slowromancap\romannumeral #1@}
\makeatother

\newcommand{\beq}{\begin{equation}}
\newcommand{\eeq}{\end{equation}}
\newcommand{\bea}{\begin{eqnarray}}
\newcommand{\eea}{\end{eqnarray}}

\renewcommand{\title}[1]{\vbox{\center\LARGE{#1}}\vspace{5mm}}
\renewcommand{\author}[1]{\vbox{\center\large{#1}}\vspace{5mm}}

\newcommand{\Tr}{\mathrm{Tr}}

\newcommand{\Cset}{{\,\,{{{^{_{\pmb{\mid}}}}\kern-.47em{\mathrm C}}}}}

\newcommand{\comment}[1]{}

\numberwithin{equation}{section}

\begin{document}

\begin{flushright}\footnotesize
\texttt{NORDITA-2014-40} \\
\texttt{UUITP-03/14}\\
\texttt{QMUL-PH-14-08}
\vspace{0.6cm}
\end{flushright}

\renewcommand{\thefootnote}{\fnsymbol{footnote}}
\setcounter{footnote}{0}

\begin{center}
{\Large\textbf{\mathversion{bold} Holographic Dual of the Eguchi-Kawai Mechanism}
\par}

\vspace{0.8cm}

\textrm{Donovan~Young$^{1}$ and
Konstantin~Zarembo$^{2,3}$\footnote{Also at ITEP, Moscow, Russia.}}
\vspace{4mm}

\textit{${}^1$ 
Centre for Research in String Theory, 
School of Physics and Astronomy,\\
Queen Mary, University of London,
Mile End Road, London, E1 4NS, UK}\\
\textit{${}^2$Nordita, KTH Royal Institute of Technology and Stockholm University,
Roslagstullsbacken 23, SE-106 91 Stockholm, Sweden}\\
\textit{${}^3$Department of Physics and Astronomy, Uppsala University\\
SE-751 08 Uppsala, Sweden}\\
\vspace{0.2cm}
\texttt{d.young@qmul.ac.uk, zarembo@nordita.org}

\vspace{3mm}


\par\vspace{1cm}

\textbf{Abstract} \vspace{3mm}

\begin{minipage}{13cm}

The holographic dual  of $\mathcal{N}=2^*$, $D=4$ supersymmetric Yang-Mills theory has many features common to 5d CFT. We interpret this as a manifestation of Eguchi-Kawai mechanism.

\end{minipage}

\end{center}

\vspace{0.5cm}



\setcounter{page}{1}
\renewcommand{\thefootnote}{\arabic{footnote}}
\setcounter{footnote}{0}

\section{Introduction}

The holographic gauge-string duality has given us  insights into many strong coupling problems in quantum field theory.
Exploring holography outside the regime of conformal
symmetry, and in situations with reduced supersymmetry, is especially interesting in this respect. Perhaps the simplest model that extends the realm of the AdS/CFT duality beyond conformal theories is ${\cal N}=2^*$ supersymmetric Yang-Mills theory (SYM), where the adjoint hypermultiplet of ${\cal
  N}=4$ SYM is given a mass, breaking both half of the supersymmetry and the
conformal invariance of the latter theory. 

The Pilch-Warner (PW) solution of type IIB supergravity \cite{Pilch:2000ue}   has long been known as the
holographic dual of ${\cal N}=2^*$ SYM  \cite{Gubser:2000nd,Brandhuber:2000ct}. The PW background consists of a domain wall that separates an asymptotically $AdS_5\times S^5$ geometry near the boundary from the near-horizon region far in the IR. The coordinate distance between the domain wall and the boundary sets the mass scale of the dual field theory. 

The $\mathcal{N}=2^*$ theory can be pictured as a flow that starts with $\mathcal{N}=4$ SYM in the UV.  At weak coupling, the IR end of the flow is pure $\mathcal{N}=2$ SYM, obtained upon integrating out the hypermultiplet. An interesting question is what happens in the IR when the coupling is not small. When the coupling is big this question can be addressed holographically. The answer is somewhat unexpected -- it turns out that the far IR regime of $\mathcal{N}=2^*$ SYM exhibits features characteristic of a five-dimensional CFT, as first observed in \cite{HoyosBadajoz:2010td}. Indeed,  the holographic entropy density of $\mathcal{N}=2^*$ SYM scales as $s\sim T^4$ at low temperatures and the speed of sound approaches $c_s^2= 1/4$, the behaviour compatible with five-dimensional scale invariance. To this list we can add the static potential that grows at large distances as $-1/L^2$  \cite{Brandhuber:1999jr,Brandhuber:2000ct}, a Coulomb law in $(4+1)$ dimensions.  As explained in \cite{HoyosBadajoz:2010td}, these observations have a geometric origin, as the far-IR geometry of the PW solution can be brought to the asymptotically $AdS_6$ form by a coordinate transformation. 

What is the origin of the fifth dimension in the dual $\mathcal{N}=2^*$ field theory? Is it possible to explain the occurrence of an extra dimension without holography? We are going to argue that the dimensional crossover is a manifestation of the Eguchi-Kawai mechanism \cite{Eguchi:1982nm,Bhanot:1982sh,Parisi:1982gp,Gross:1982at} at large-$N$ and large 't~Hooft coupling (a similar explanation was given in \cite{HoyosBadajoz:2010td}). Another holographic realization of the Eguchi-Kawai reduction is the thermal AdS dual of  $\mathcal{N}=4$ super-Yang-Mills theory at finite temperature \cite{Furuuchi:2005qm,Furuuchi:2005eu,Poppitz:2010bt,Unsal:2010qh}.

The Eguchi-Kawai reduction arises in large-$N$ gauge theories with flat directions in the potential. For the mechanism to work, translation symmetry along flat directions should not be broken by quantum fluctuations. Most commonly, the Eguchi-Kawai mechanism relies on fields with commutator couplings.
Suppose that fields $\Phi _I$, in the adjoint of the gauge group $U(N)$, enter the Lagrangian only through couplings of the form $\mathop{\mathrm{tr}}[\Phi _I,\phi ]^2$. The potential then has flat directions along the diagonal components of $\Phi _I$, which can thus condense. In the original Eguchi-Kawai construction, the fields $\Phi _I$ are the gauge potentials $A_\mu $ of the reduced theory in zero space-time dimensions \cite{Eguchi:1982nm}. The field that condenses in $\mathcal{N}=2^*$ SYM is the scalar from the vector multiplet, and we will suppress the index $I$ in what follows:
\begin{equation}\label{vevPhi}
 \left\langle \Phi \right\rangle =\mathop{\mathrm{diag}}\left(a_1,\ldots ,a_N\right).
\end{equation}

In the presence of the condensate, the field components with color indices $i,j$ acquire masses $m^2_{ij}=m^2+(a_i-a_j)^2$, where $m$ is the bare mass in the Lagrangian and $(a_i-a_j)^2$ is the Higgs mass that comes from the commutator term. In the particular case of $\mathcal{N}=2^*$ SYM, $m=0$ for the vector multiplet and $m=M$ for the matter hypermultiplet, where $M$ is the mass scale in the Lagrangian.

\begin{figure}[t]
\begin{center}
 \centerline{\includegraphics[width=8cm]{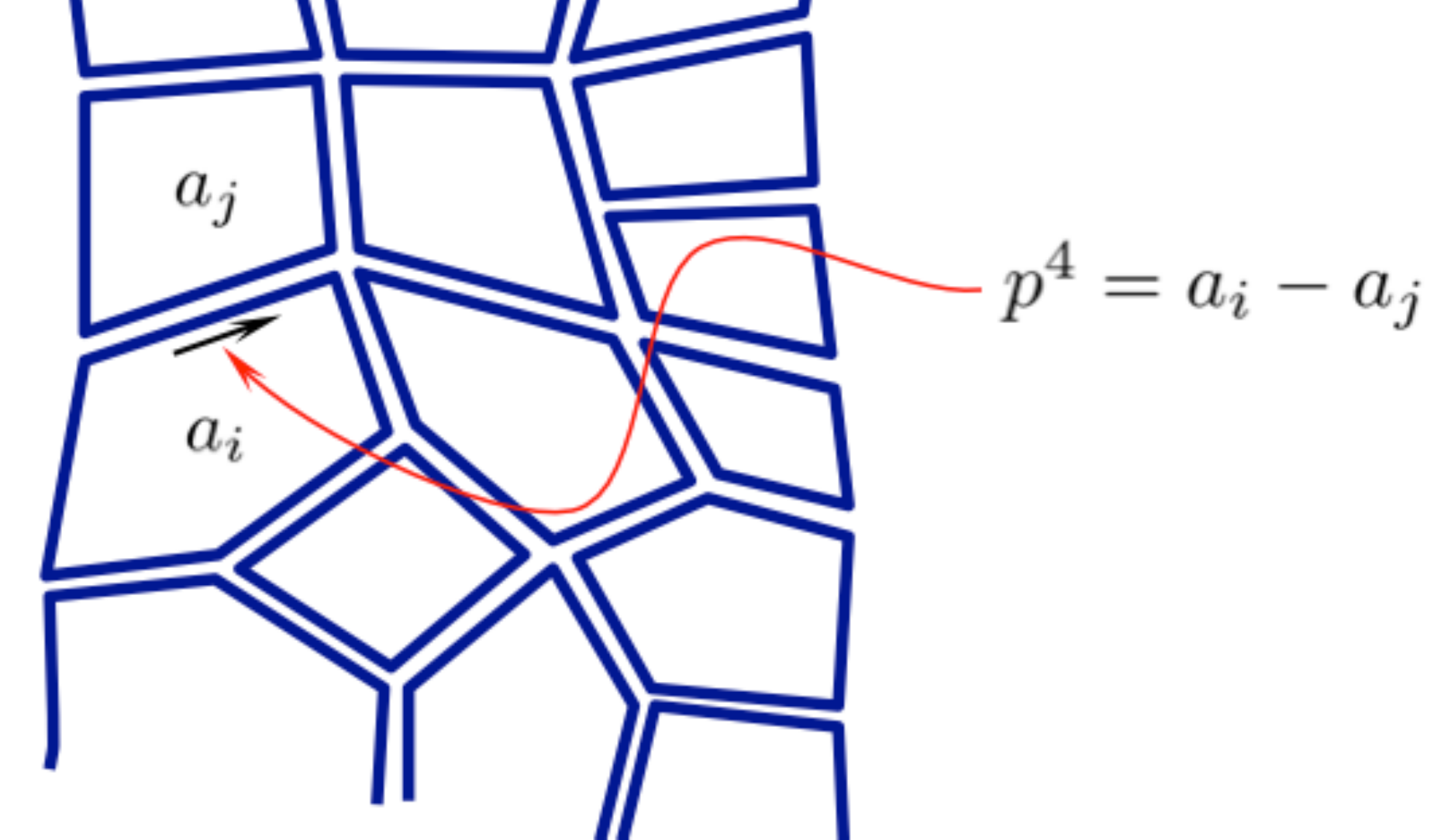}}
\caption{\label{Pdiag}\small Each facet of a planar diagram is associated with a color index. The line that separates facets with indices $i$ and $j$ is assigned the momentum $p^4=a_i-a_j$. The map $\{a_i\}\rightarrow \{p^4_l\}$ is one-to-one for a planar diagram.}
\end{center}
\end{figure}

The propagators in the diagonal background (\ref{vevPhi}) are of the form:
\begin{equation}
 \wick[u]{1}{<1\phi ^i_j(p)>1\phi ^k_l(-p)}
 =\frac{\delta ^i_l\delta ^k_j}{p^2+\left(a_i-a_j\right)^2+m^2}\,.
\end{equation}
Each index loop of a Feynman diagram in 't~Hooft's double-line notation carries a particular color and hence a particular $a_i$. Summation over color indices then amounts to averaging over $a_i$'s with the weight
\begin{equation}\label{eigenvalue-distribution}
 \rho (a)=\frac{1}{N}\,\sum_{i=1}^{N}\delta \left(a-a_i\right).
\end{equation}
In any planar diagram, the number of independent $a_i$'s is the same
as the number of loop momenta, up to an overall shift $a_i\rightarrow
a_i+\,{\rm const}\,$. The eigenvalue density of the symmetry-breaking
VEV becomes a smooth function in the large-$N$ limit, and the
difference $a_i-a_j$ can be interpreted as the fifth component $p^4$ of the momentum flowing through the $ij$ line (fig.~\ref{Pdiag}). Of course, color averaging resembles momentum integration only if the distribution $\rho (x)$ is flat within a sufficiently wide interval $[-\Lambda ,\Lambda ]$, where $\Lambda $ then plays the r\^ole of a UV cutoff.
It can be shown that color averaging is equivalent to momentum integration to any order of planar perturbation theory \cite{Gross:1982at}, provided that the distribution $\rho (a)$ is sufficiently flat.

A well-known obstacle to Eguchi-Kawai reduction is that $\rho (a)$, in principle a dynamical quantity, is not really flat \cite{Bhanot:1982sh}.  For the reduction to work the eigenvalues $a_i$ must be distributed more or less
uniformly over a fairly large interval, larger than any physical mass scale in the problem. This is not the case in the simplest Eguchi-Kawai model, where the eigenvalues tend to clump around zero at weak coupling \cite{Bhanot:1982sh}. The problem can be circumvented  in models with adjoint matter and/or double-trace couplings \cite{Kovtun:2007py,Unsal:2008ch}.

In the case at hand, the dimensional crossover for sure does not happen at weak coupling, when  $\mathcal{N}=2^*$ SYM  flows to the pure gauge $\mathcal{N}=2$ theory in the IR. 
The strong-coupling nature of the gravitational description is consequently crucial for opening up of the fifth dimension, observed in the holographic dual of  $\mathcal{N}=2^*$ SYM. The difference must be accounted for by the structure of the eigenvalue density (\ref{vevPhi}). 

The eigenvalue density in $\mathcal{N}=2^*$ SYM is actually known. At strong coupling, it can be computed holographically by the probe analysis of the PW background \cite{Buchel:2000cn}:
\begin{equation}\label{strong-eigen}
 \r(a) =\frac{2}{\pi\m^2}
\sqrt{\m^2-a^2},\quad\m = \frac{\sqrt{\l}M}{2\pi},
\end{equation}
where $\lambda =g_{\rm YM}^2N$ is the 't~Hooft coupling.
The same eigenvalue density can be calculated directly from field theory, without any reference to holography, by localizing the path integral on $S^4$ \cite{Pestun:2007rz} and taking the large-$N$ limit of the localization partition function \cite{Buchel:2013id}. The result (\ref{strong-eigen}) arises in the strong-coupling limit, $\lambda \rightarrow \infty $. The exact eigenvalue distribution   \cite{Russo:2013qaa,Russo:2013kea,Russo:2013sba} flattens with growing $\lambda $, as already evident from the strong-coupling expression (\ref{strong-eigen}). The eigenvalue cutoff, $\mu= \sqrt{\l}M/2\pi$, becomes parametrically larger than the mass scale $M$ if $\lambda \gg 1$, and we may then expect the Eguchi-Kawai mechanism to work. 

In sec.~\ref{sec:potential} we provide further evidence for
Eguchi-Kawai mechanism in strongly-coupled $\mathcal{N}=2^*$ SYM by
revisiting the computation of the static potential, as measured by the infinite rectangular Maldacena-Wilson
loop. For
small quark-anti quark separations $L$, one finds the expected $1/L$
Coulombic behaviour encountered in the ${\cal N}=4$ SYM / $AdS_5\times
S^5$ duality \cite{Maldacena:1998im,Rey:1998ik}. For $L$ much greater than the (inverse of the) ${\cal
  N}=2^*$ mass scale $M$, the potential morphs to that expected from a
5-dimensional gauge theory, i.e. it goes as $1/L^2$
\cite{Brandhuber:2000ct}. This behaviour has been understood as due to
the continuous distribution of branes found at the enhan\c{c}on radius
\cite{Freedman:1999gk}. Here we relate this behaviour with the Eguchi-Kawai mechanism.
 
We will provide insight into the mechanism responsible for the
$1/L^2$ potential by considering simple one-loop diagrams in the gauge theory
at weak coupling, albeit inserting the strong-coupling form of the VEV
eigenvalue distribution (\ref{strong-eigen}). 
We find that a familiar phenomenon is at
play: the condensed eigenvalues play the role of an extra-dimensional
momentum, which when integrated over produces the $1/L^2$ form of the
potential.

The simplest relevant deformation of the standard AdS/CFT setup, $\mathcal{N}=2^*$ holography, was studied from many points of view.
In particular, D-brane probes have provided a wealth of information about the
PW geometry
\cite{Buchel:2000cn,Evans:2000ct,Babington:2001nw,Evans:2005ti,Albash:2011nw}.
The holographic dual of $\mathcal{N}=2^*$ SYM at finite temperature was constructed in  \cite{Buchel:2003ah}, opening the avenue for studying thermodynamics of this theory at strong coupling \cite{Buchel:2004hw,Benincasa:2005iv,Buchel:2007vy,Buchel:2008uu,HoyosBadajoz:2009pv,Buchel:2010ys,Hoyos:2011uh,Albash:2011dq,Buchel:2012gw}. It is this analysis 
 that first pointed to the 5d nature of the strongly-coupled $\mathcal{N}=2^*$ SYM \cite{HoyosBadajoz:2010td}. In addition, classical string probes  \cite{Dimov:2003bh}, supersymmetry properties  \cite{Pilch:2003jg},  and  entanglement
entropy \cite{Lewkowycz:2012qr} of the PW solution have been investigated. 
Owing to the ability to apply localization to the theory on $S^4$
\cite{Pestun:2007rz}, the strong-coupling behaviour of the free
energy, Wilson loops \cite{Buchel:2013id,Russo:2012kj}, and the
theory's phases \cite{Russo:2013qaa,Russo:2013kea,Russo:2013sba} have been explored
using matrix model techniques. Comparisons using supergravity
computations on the Pilch-Warner background (and its generalization
for the case of $S^4$ boundary  \cite{Bobev:2013cja,Balasubramanian:2013esa})
have been successfully made for Wilson loops \cite{Buchel:2013id} and
recently for the free energy \cite{Bobev:2013cja}.

To further study the IR CFT of the strongly coupled $\mathcal{N}=2^*$ SYM, in section \ref{sec:2pt} we consider the boundary-to-boundary
propagator for massive scalar fields on the Pilch-Warner background,
for separations $\gg M^{-1}$.

\section{Static potential}\label{sec:potential}

\begin{figure}[t]
\begin{center}
 \centerline{\includegraphics[width=6cm]{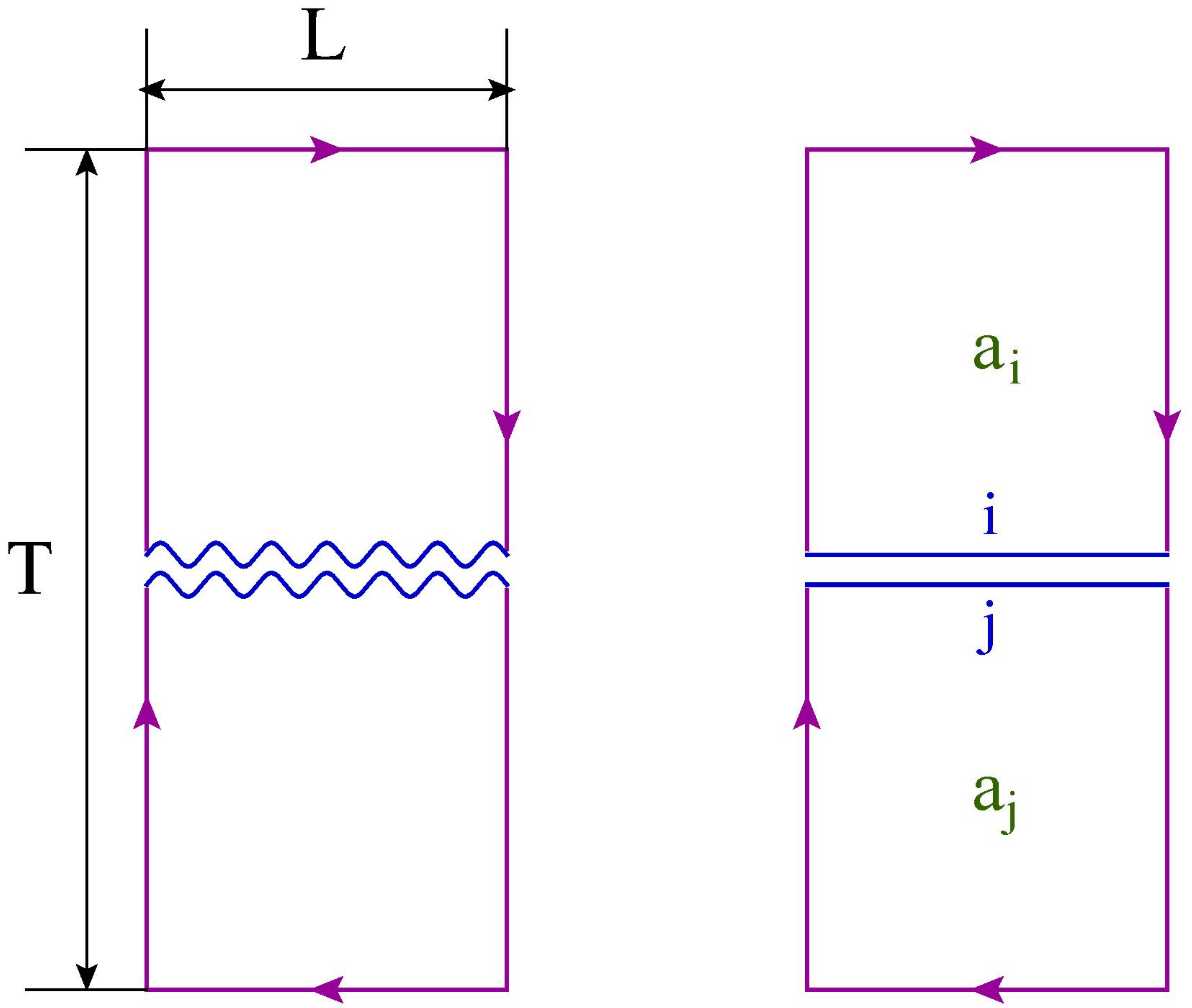}}
\caption{\label{1llp}\small The rectangular contour used to compute the static
  potential. At the leading order in perturbation theory the potential is given by the exchange of
  $\varphi$ and $A_\mu $ fields between opposite sides of the Wilson loop. Each line carries color indices $i$ and $j$ and associated eigenvalues of the scalar VEV $a_i$ and $a_j$.}
\end{center}
\end{figure}

The static potential is defined through the expectation value of the rectangular Wilson loop, which in an $\mathcal{N}=2$ supersymmetric theory couples also to scalars.
The ${\cal N}=2^*$ theory has two real scalars in the vector multiplet. 
Here we label them $\Phi$ and $\Phi'$. The
Maldacena-Wilson loop is taken to couple to just one of those, the same one that takes on the vacuum expectation value (\ref{vevPhi}):
\be \label{what Wilson loop}
W (C) = \left\langle\frac{1}{N}\Tr\, {\cal P}\exp \oint_C d\t \,\Bigl( i\dot x^\m(\t) A_\m + 
|\dot x(\t)| \Phi \Bigr)\right\rangle.
\ee
The potential is computed according to the identification
\be
 V(L)=-\lim_{T\rightarrow \infty }\frac{1}{T}\,\ln W(C_{T\times L})
\ee
for a rectangular contour with sides $L$ and $T$ as shown in fig.~\ref{1llp}. 

The static potential $V(L)$ at strong coupling is given by the action
of a string worldsheet with a boundary consisting of two infinite
anti-parallel lines separated by a distance $L$. At small $L$, the potential has the expected $1/L$ behaviour at short distances while at large distances it approaches a constant with the deviation scaling as $1/L^2$. We shortly review the supergravity calculation of the static potential, which can be found in appendix A of \cite{Brandhuber:1999jr}, and then demonstrate that very similar behaviour arises on the field-theory side if we use the following heuristic prescription. The potential depends on the eignvalue distribution of the symmetry-breaking condensate. If the strong-coupling distribution (\ref{strong-eigen})
 is inserted in the tree-level exchange diagram, then at large distances the potential has a $1/L^2$ fall-off,  as a consequence of the Eguchi-Kawai mechanism.

\subsection{Strong coupling}

The PW background is a five-dimensional domain wall whose  metric is also warped with respect to the coordinates of the internal manifold, a deformed $S^5$. The five-dimensional metric therefore depends on where on the internal manifold the string sits. This, in turn, is determined by the scalar couplings of the Wilson loop. 
The deformation of $S^5$ preserves the $S^1\times S^3$ foliation, with $S^1$ roughly speaking dual to the vector multiplet and $S^3$ dual to the hypermultiplet scalars. The Wilson loop (\ref{what Wilson loop}) only couples to the vector multiplet scalar $\Phi $, so the dual string worldsheet sits at the locus where $S^3$ shrinks to zero size, $\T=\pi/2$ in the notation of \cite{Pilch:2000ue}, and since there is no coupling to $\Phi '$, at $\varphi =0$, where $\varphi $ is the coordinate on $S^1$. 

The five-dimensional PW metric, restricted to $\T=\pi/2$, $\varphi =0$ and  
transformed to the string frame, is given by
\be\label{slicePW}
ds^2 = \frac{A}{c^2-1}\, M^2 dx_\m^2 +
\frac{1}{A(c^2-1)^2}\,dc^2,
\ee
where\footnote{The $\rho ^6$ used in
  \cite{Pilch:2000ue} is here denoted by $A$.}
\be\label{Adefinition}
A = c + \frac{c^2-1}{2}\log\frac{c-1}{c+1}.
\ee
The boundary is at $c=1$, with $c$ related to the conventional radial coordinate $z$ of $AdS_5$ as
\begin{equation}\label{c-to-z}
 c=1+\frac{z^2M^2}{2}+\ldots 
\end{equation}
For the minimal surface parameterized by $t$ and $c=c(x)$, the Lagrangian density of the Nambu-Goto action is
\be
{\cal L} = \frac{M}{c^2-1}\sqrt{
A^2M^2+\frac{
(c')^2
}{c^2-1}},
\ee
where $'$ denotes $d/dx$. 
Going through the usual steps we obtain the potential in parametric form:
\bsp\label{sys}
&\frac{ML}{2} = \int_1^{c_m}  \frac{dc}{A\sqrt{c^2-1}}
\frac{1}{\sqrt{
\frac{(c_m^2-1)^2A^2}{(c^2-1)^2A^2_m} -1}},\\
&V(L) = \frac{M\sqrt{\l}}{\pi}\int_{1+\frac{\e^2M^2}{2}}^{c_m}\frac{dc}{(c^2-1)^{3/2}}
\,\,\frac{1}{\sqrt{
1
-\frac{A^2_m}{A^2}\frac{(c^2-1)^2}{(c_m^2-1)^2}}}
-\frac{\sqrt{\lambda }}{\pi \epsilon }
\\
&\hphantom{V(L)}=\frac{M\sqrt{\l}}{\pi}\left\{
 \int_{1}^{c_m}\frac{dc}{(c^2-1)^{3/2}}
 \left[
\frac{1}{\sqrt{
1
-\frac{A^2_m}{A^2}\frac{(c^2-1)^2}{(c_m^2-1)^2}}}-1\right]-\frac{c_m}{\sqrt{c_m^2-1}}\right\}.
\end{split}
\ee
where $c_m$ is the value of the coordinate $c$ at the midpoint where the
string reaches its maximum depth in the holographic direction, and $A_m = A(c_m)$. The overall factor of $\sqrt{\lambda }/2\pi $ is the dimensionless string tension.
We employ the AdS-based prescription for regularization of the boundary divergence, which guarantees the match to the $\mathcal{N}=4$ result \cite{Maldacena:1998im,Rey:1998ik} at short distances:
\begin{equation}
 V(L)=-\frac{M}{\Gamma ^2\left(\frac{1}{4}\right)}\,\sqrt{\frac{\lambda \pi }{c_m-1}}+\ldots 
 =
 -\frac{4\pi ^2\sqrt{\lambda }}{\Gamma ^4\left(\frac{1}{4}\right)}\,\,\frac{1}{L}+\ldots \qquad \left(LM\ll 1\right).
\end{equation}

In the opposite limit of $LM \gg 1$ the first integral in
(\ref{sys}) is dominated by large $c$. This is because the string
worldsheet quickly drops into the bulk, whereupon it turns over and
stays at essentially constant $c=c_m \propto L$ before turning back to the
boundary in a symmetric fashion. Then \cite{Brandhuber:1999jr}
\be\label{strong}
V(L)=-\frac{M\sqrt{\lambda }}{\pi }\left(
1+\frac{\sqrt{\pi }\Gamma \left(\frac{2}{3}\right)}{2\Gamma \left(\frac{1}{6}\right)c_m^2}+\ldots 
\right)
=
-\frac{M\sqrt{\lambda }}{\pi }-\frac{9\Gamma ^3\left(\frac{2}{3}\right)\sqrt{\pi  }}{2\Gamma ^3\left(\frac{1}{6}\right)}\,\,\frac{\sqrt{\lambda }}{ML^2}+\ldots \qquad \left(LM\gg 1\right), 
\ee
and so, after the perimeter law, the potential goes as $1/L^2$ at
large-$L$. The constant term in the IR asymptotics of the potential reflects the finite self-energy of well-separated quarks. The self-energy can be computed exactly  \cite{Buchel:2013id,Russo:2013qaa} by a first-principles field-theory calculation (from localization \cite{Pestun:2007rz}). The strong-coupling interpolation of this result perfectly matches with the first term in (\ref{strong})  \cite{Buchel:2013id}. The potential is partially screened at large distances, and behaves as if the theory were five-dimensional which is another manifestation of the phenomenon observed in \cite{HoyosBadajoz:2010td}, that the holographic dual of $\mathcal{N}=2^*$ SYM flows to a 5d CFT in the IR.

\begin{figure}[t]
\begin{center}
 \centerline{\includegraphics[width=7cm]{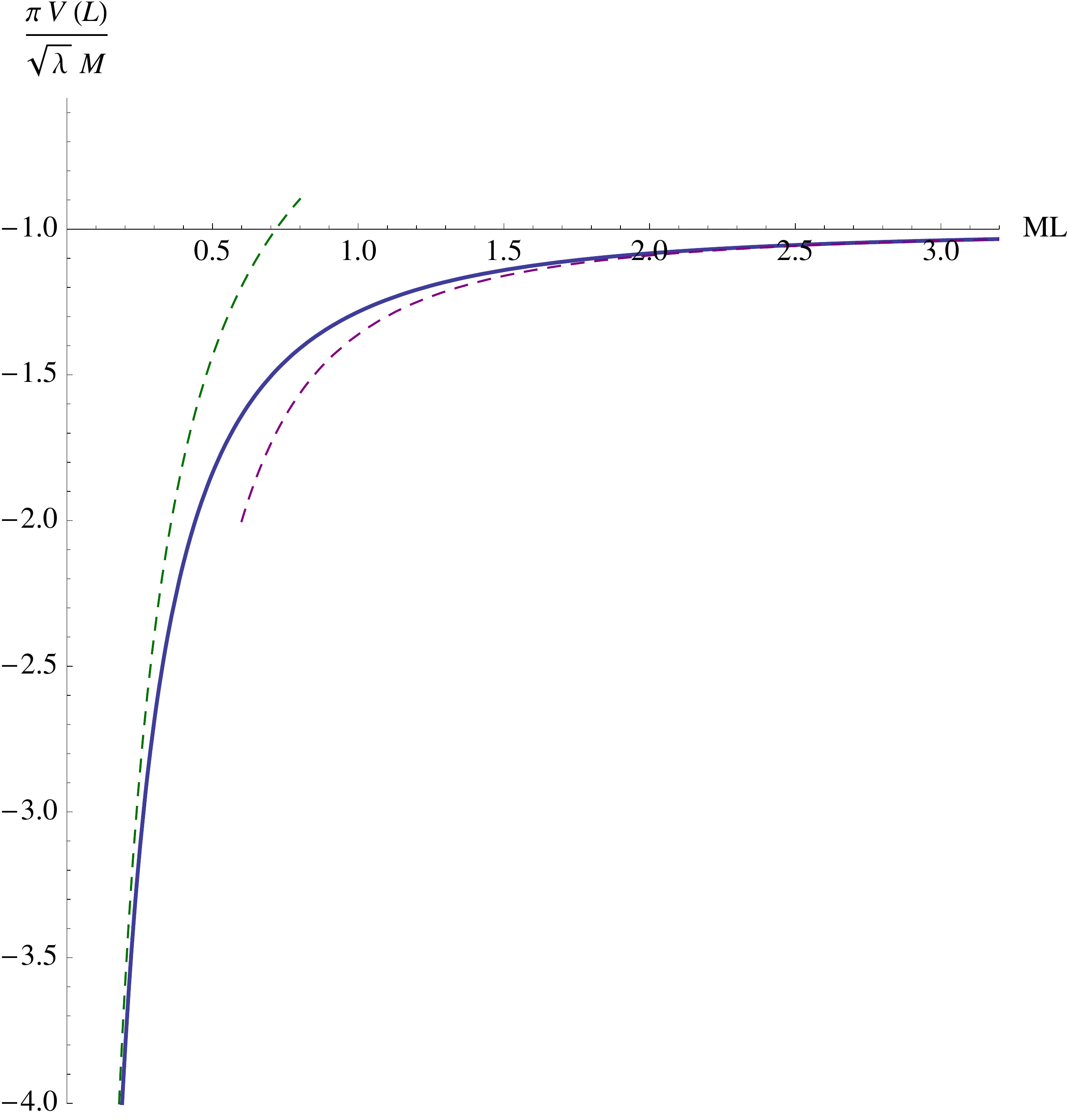}}
\caption{\label{fig:potstrong}\small The static potential at strong coupling.  Asymptotic behaviour at small and large distances is shown in dashed lines.}
\end{center}
\end{figure}

The full strong-coupling potential is shown in fig.~\ref{fig:potstrong}. The transition from the Coulomb law at short distances to the partially screened behaviour at large distances happens, as expected, at $L\sim 1/M$.

\subsection{Weak coupling}

The leading order at weak coupling is the one-particle exchange diagrams in fig.~\ref{1llp}, where the static potential is given by the spatial Fourier transform of the propagator at $p^0=0$ (subtracting the constant self-energy):
\be\label{weak}
V(L)  =- \frac{\l}{N^2}\sum_{i,j=1}^N \int \frac{d^3p}{(2\pi)^3} \frac{e^{i\vec p
    \cdot \vec L}}{\vec p^2 + a_{ij}^2}\,.
\ee
The potential depends on the eigenvalue distribution (\ref{eigenvalue-distribution}). Let us now us plug-in the strong-coupling eigenvalue density (\ref{strong-eigen}) in the simple tree-level expression for the potential. This is clearly not a consistent procedure, but it does illustrate how the Eguchi-Kawai mechanism works in this context. 
We have:
\be
V(L) = -\l \int \frac{d^3p}{(2\pi)^3} \,e^{i\vec p
    \cdot \vec L}\int
da\int db\,\frac{\r(a)\r(b)}{\vec p^2 + (a-b)^2}.
\ee
If we are interested in $V(L)$ for large values of $L\gg M^{-1}$, then the
integration over $\vec p$ is dominated by $p \sim L^{-1}$. But then the
integration over the eigenvalues is dominated by the region $a-b\sim
p\sim L^{-1}$. Given the distribution $\r(a)$ at strong coupling given
in (\ref{strong-eigen}), we see that $a$ and $b$ themselves are
${\mathcal O}(\sqrt{\l}M)$. Since we are taking $M\gg L^{-1}$ we see that $a$ and
$b$ are individually much larger then their separation $a-b$. We may
thus define a ``fourth momentum'' $p_4 \equiv a-b$ and write
\bsp
V(L\gg M^{-1}) &\simeq - 2\pi\l \int_{-\m}^\m da \, \r^2(a)\times
\int \frac{d^3p}{(2\pi)^3}
\int \frac{dp_4}{2\pi}\,
 \,\frac{e^{i\vec p \cdot \vec L}}{\vec p^2+p_4^2} \\
&=-\frac{32\l}{3\pi\m} \times \frac{1}{4\pi^2 L^2}
=-\frac{16}{3\pi^2} \frac{\sqrt{\l}}{ML^2}.
\end{split}
\ee 
The parametric dependence on both $L$ and $\l$ is in agreement with the strong-coupling result (\ref{strong}). The numerical prefactors in front of $\sqrt{\lambda }/ML^2$ are quite different, which is not unexpected as the weak-coupling calculation was not really self-consistent. We believe however, that the qualitative reason for the change from the  $1/L$ Coulomb behaviour to $1/L^2$  is a manifestation of the same mechanism in both cases. At higher orders of perturbation theory, which have to be taken into account if $\lambda $ is big, the difference $a_i-a_j$ across any line of a planar diagram effectively plays the r\^ole of the fifth momentum, while the wide spread of the eigenvalues at strong coupling supplies the flat measure for momentum integration. It would be interesting to make this qualitative picture more precise. 

\section{Two-point function}
\label{sec:2pt}

To further study the effective IR CFT of strongly-coupled
$\mathcal{N}=2^*$ theory, we would like to compute holographic
two-point functions in the Pilch-Warner background. This is a
formidable task, given the complexity of the supergravity
solution. But since we are interested in qualitative features of the
effective CFT and thus focus on the IR behaviour, we will study a
simpler problem by truncating the geometry to its 5d $AdS$-like slice
(\ref{slicePW}), keeping in mind the Kaluza-Klein-type decomposition
of the full 10d field. We will study a scalar field in this geometry
minimally coupled to the metric either in the string frame, as written
in (\ref{slicePW}), or in the Einstein frame where the metric is
multiplied the dilaton factor:
\begin{equation}
 \,{\rm e}\,^{-\frac{\phi }{2}}=\sqrt{c}.
\end{equation}
Normally, the large-distance behaviour of holographic two-point functions is governed by the geodesic approximation. This is true in $AdS_5$ and in many of its non-conformal cousins, but here the geodesic approximation does not work, for the following reason: the (renormalized) length of the geodesic that connects two points $x$ and $y$ on the boundary grows with $|x-y|$, but not sufficiently fast. The geodesic distance saturates at large coordinate separation and goes to a constant as $|x-y|\rightarrow \infty $. In this sense all the points on the boundary are close to each other; the geodesic distance never becomes large and the WKB approximation never becomes accurate for the boundary-to-boundary propagator. We thus need to study the Klein-Gordon equation directly.

We start with the five-dimensional metric 
\be\label{fh}
ds^2 = f(c)\, dx^\m dx_\m + h(c) \,dc^2.
\ee
The functions $f(c)$ and $h(c)$ are defined in (\ref{slicePW}), (\ref{Adefinition}):
\begin{equation}
 f(c)=\frac{M^2Ac^\nu }{c^2-1}\,,\qquad h(c)=\frac{c^\nu }{A\left(c^2-1\right)^2}\,,
\end{equation}
and 
\begin{equation}
 \nu =
\begin{cases}
 0 & {\rm in~string~frame}
\\
 \frac{1}{2} & {\rm in~Einstein~frame}.
\end{cases}
\end{equation}
We leave $f$ and $h$ unspecified 
for the time being. 

The Green's function $G_\Delta (c,x^\m;c',{x'})$ for a
scalar field dual to an operator with UV dimension $\Delta $ satisfies
\be\label{KG-eq}
\left[ -\frac{1}{f^2\sqrt{h}} \,\p_c \,\frac{f^2}{\sqrt{h}}\,
  \p_c - \frac{1}{f}\,\p_\m\p^\m +\Delta \left(\Delta -4\right)\right] G_\Delta (c,x;c',{x'}) 
= \frac{\d^4(x-x')\d(c-c')}{f^2\sqrt{h}}\,.
\ee
The two-point function is given by the boundary-to-boundary limit of the bulk propagator:
\begin{equation}
 \left\langle \mathcal{O}(x)\mathcal{O}(x')\right\rangle
 =\frac{2^{1-\Delta }\pi ^2M^{2\Delta }}{\Delta -1}\,\lim_{\varepsilon \rightarrow 0}
 \varepsilon ^{-\Delta }G_\Delta \left(1+\varepsilon ,x;1+\varepsilon ,x'\right).
\end{equation}
The factors of $2$ and $M$ reflect the relationship (\ref{c-to-z}) between the coordinate $c$ of the PW metric and the canonical radial coordinate of $AdS_5$.

The Klein-Gordon equation can be brought to the Schr\"odinger form by multiplying the Green's function with  $fh^{-1/4}$. The resulting Schr\"odinger equations reads
\be\label{U}
-U_\a'' + \left[
 \frac{\left(fh^{-\frac{1}{4}}\right)''}{fh^{-\frac{1}{4}}}
 +\Delta \left(\Delta -4\right)h
 -\alpha ^2\,\frac{h}{f}
\right] U_\a = 0,
\ee
where $'$ denotes $d/dc$. Since the effective potential starts with an infinite wall\footnote{Or an infinite dip, depending on $\Delta $.} at $c\rightarrow 1$ and goes to zero sufficiently fast at infinity, there is one eigenfunction for each $\alpha >0$. The eigenfunctions can be normalized as
\begin{equation}\label{norm}
 \int_{1}^{\infty }dc\,\,\frac{h(c)}{f(c)}\,U^*_{\alpha'} (c)U_\alpha(c)=\delta \left(\alpha '{}^2-\alpha ^2\right),
\end{equation}
and as a consequence satisfy the completeness condition
\begin{equation}\label{completeness-U}
 \int_{0}^{\infty }d\alpha^2 \,U_\alpha (c)U_{\alpha }^*(c')=\frac{f(c)}{h(c)}\,\delta \left(c-c'\right).
\end{equation}
The bulk-to-bulk propagator is expressed in terms of the eigenfunctions $U_\alpha (c)$ as
\begin{equation}
 G_\Delta (c,x;c',x')=\frac{h^{\frac{1}{4}}(c)h^{\frac{1}{4}}(c')}{f(c)f(c')}
 \int_{}^{}\frac{d^4p}{\left(2\pi \right)^4}\,\,\,{\rm e}\,^{ip(x-x')}
 \int_{0}^{\infty }d\alpha ^2\,\,\frac{U_\alpha (c)U_\alpha^* (c')}{p^2+\alpha ^2}\,.
\end{equation}
This can be verified rather directly by applying the Klein-Gordon operator (\ref{KG-eq}) to $G_\Delta $ so defined and using the properties (\ref{U})--(\ref{completeness-U}) above. 

The spectral decomposition of the bulk propagator yields the K\"{a}ll\'{e}n-Lehmann representation for the two-point function:
\begin{equation}
 \left\langle \mathcal{O}(x)\mathcal{O}(x')\right\rangle=
 \int_{0}^{\infty }d\alpha^2 \,\rho _\Delta (\alpha )
 \int_{}^{}\frac{d^4p}{\left(2\pi \right)^4}\,\,\,{\rm e}\,^{ip(x-x')}
 \,\frac{1}{p^2+\alpha ^2}\,,
\end{equation}
with the spectral weight
\begin{equation}\label{spectral-rho}
 \rho _\Delta (\alpha )=\frac{2^{2-\Delta }\pi ^2M^{2\Delta -4}}{\Delta -1}\,\lim_{\varepsilon \rightarrow 0}\varepsilon ^{1-\Delta }\left|U_\alpha (1+\varepsilon )\right|^2.
\end{equation}

This is the standard holographic machinery. We need to specify it for the case of the PW background.
The PW metric behaves near the boundary as
\begin{equation}
 f(c)\simeq \frac{M^2}{2\left(c-1\right)}\,,\qquad 
 h(c)\simeq \frac{1}{4\left(c-1\right)^2}
 \qquad \qquad \left(c\rightarrow 1\right).
\end{equation}
Upon the change of variables (\ref{c-to-z}) the PW metric in this
approximation reduces to the standard (dimensionless) metric of the Poincar\'{e} patch of $AdS_5$. In the deep-IR
where $c \gg 1$, we have instead
\be
f(c) \simeq \frac{2M^2}{3c^{3-\nu }},\qquad
h(c) \simeq \frac{3}{2 c^{3-\nu }}\qquad \qquad \left(c\rightarrow \infty \right).
\ee
In either case, (\ref{U}) reduces to the Bessel equation:
\begin{equation}
 -U_\a'' + \left[\frac{\Delta \left(\Delta -4\right)+3}{4\left(c-1\right)^2}-\frac{\alpha ^2}{2M^2(c-1)}\right]U_\a = 0\qquad \qquad \left(c\rightarrow 1\right),
\end{equation}
or 
\begin{equation}
  -U_\a'' +\left[ \frac{3\left(3-\nu \right)\left(13-3\nu \right)}{16c^2}
  -\frac{9\alpha ^2}{4M^2}\right]U_\a = 0\qquad \qquad \left(c\rightarrow \infty \right).
\end{equation}
We therefore find:
\begin{equation}\label{UC1}
 U_\alpha (c)\simeq C_1\sqrt{c-1}\,J_{\Delta -2}\left(\frac{\alpha}{M}\,\sqrt{2(c-1)}\right)
\qquad \qquad \left(c\rightarrow 1\right)
\end{equation}
and
\begin{equation}\label{UC2}
 U_\alpha (c)\simeq C_2\sqrt{c}\,J_{\frac{11-3\nu }{4}}\left(\frac{3\alpha c }{2M}\right)
\qquad \qquad \left(c\rightarrow \infty \right).
\end{equation}
The constants $C_1$ and $C_2$ are determined by matching the asymptotic solutions in the intermediate region $c\sim 1$ and by the overall normalization condition. To find the spectral weight of the two-point function, according to (\ref{spectral-rho}), we need to calculate one of these constants, the $C_1$.

The spectral weight at large $\alpha $, which determines the short-distance asymptotics of the two-point functions, can be inferred from (\ref{UC1}) alone. Indeed, at large $\alpha $, the eigenfunction enters the oscillating regime long before reaching the matching region of $c\sim 1$. It can be seen that the phase of oscillations changes much slower with $c$ in the region $c\ll 1$, because of the square root in the argument of the Bessel functions. This range of $c$, not very close to the boundary, but still such that $c$ is not extremely large, thus gives the leading contribution to the normalization integral, which thus determines the constant $C_1$ to be
\begin{equation}
 C_1\simeq 1\qquad \qquad \left(\alpha \rightarrow \infty \right).
\end{equation}
This yields:
\begin{equation}
 \rho _\Delta (\alpha )\simeq \frac{\pi ^2}{\Gamma \left(\Delta \right)\Gamma \left(\Delta -1\right)}\,\left(\frac{\alpha }{2}\right)^{2\Delta -4}
 \qquad \qquad \left(\alpha \rightarrow \infty \right),
\end{equation}
 the spectral weight of canonically normalized scalar of dimension $\Delta $:
\begin{equation}
 \left\langle \mathcal{O}(x)\mathcal{O}(0)\right\rangle\simeq 
 \frac{1}{|x|^{2\Delta }}\qquad \qquad \left(M|x|\ll 1 \right).
\end{equation}

\begin{figure}[t]
\begin{center}
 \centerline{\includegraphics[width=8cm]{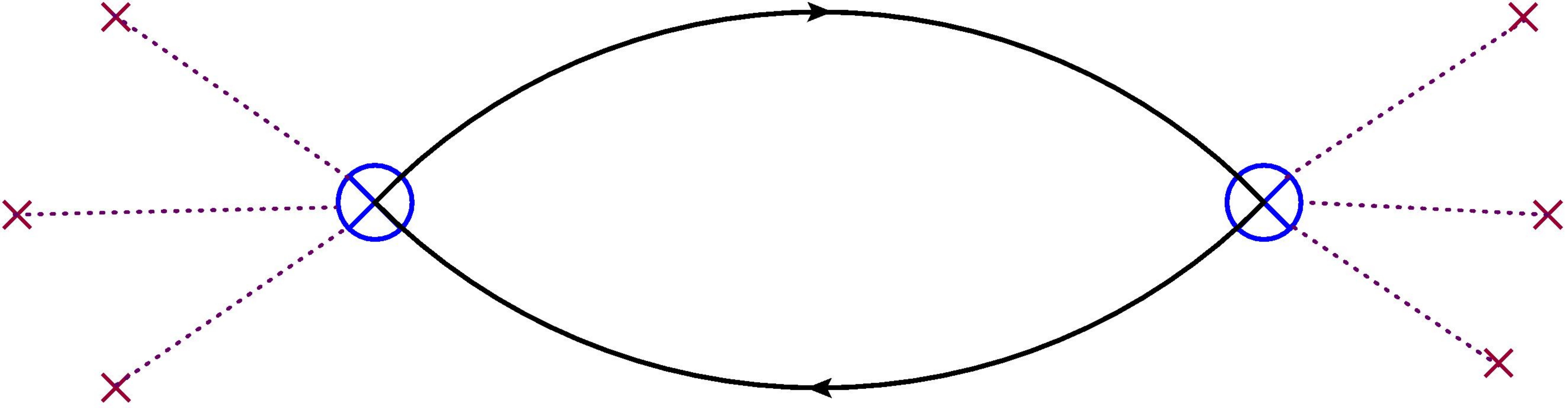}}
\caption{\label{2point-fig}\small Two-point function of the operator $\mathop{\mathrm{tr}}\Phi ^3\bar{\Psi }\Psi $ in the presence of the scalar condensate.}
\end{center}
\end{figure}

In the opposite case of small $\alpha $, the wavefunction is small near the boundary, and the large-distance asymptotics (\ref{UC2}) saturates the normalization integral. This determines the constant $C_2$:
\begin{equation}
 C_2\simeq \frac{1}{\sqrt{2}}\qquad \qquad \left(\alpha \rightarrow 0\right).
\end{equation}
Matching (\ref{UC2}) with (\ref{UC1}) at $c\sim 1$ we find:
\begin{equation}
 C_1\sim \alpha ^{\frac{19-3\nu }{4}-\Delta }\qquad \qquad \left(\alpha \rightarrow 0\right)
\end{equation}
and
\begin{equation}
 \rho _\Delta (\alpha )\sim C_1^2\alpha ^{2\Delta -4}\sim \alpha ^{\frac{11-3\nu }{2}}\qquad \qquad \left(\alpha \rightarrow 0\right).
\end{equation}
The two-point function, consequently, behaves at large distances as
\begin{equation}
 \left\langle \mathcal{O}(x)\mathcal{O}(0)\right\rangle\simeq 
 \frac{\,{\rm const}\,}{|x|^{\frac{19-3\nu }{2}}}\qquad \qquad \left(M|x|\gg 1 \right).
\end{equation}

A peculiar feature of this result is that the IR scaling dimension does not depend on the UV dimension, or in other words on the mass of the scalar  field in the bulk. It depends however on the specific coupling of the dual field to the supergravity background. For instance, the IR dimension is $\Delta _{IR}=19/4$ for the scalar with the minimal coupling in the string frame and $\Delta _{IR}=35/8$ in the Einstein frame.  Fields with the same coupling to the supergravity background, but with different masses lead to distinct scaling dimensions in the UV but the same dimension in the IR. Roughly speaking, all Kaluza-Klein modes of the same 10d field will have degenerate IR scaling dimensions. How can that happen? There is no symmetry that could explain such enormous degeneracy. Here we provide a qualitative explanation based on the weak-coupling intuition. Consider an operator that consists of $L$ scalar fields and two fermions. This operator has the UV dimension $L+3$ (at weak coupling). This follows from connecting all the fields in the two operators pairwise by propagators. But in the presence of the vacuum expectation value (\ref{vevPhi}), the scalar lines can end on the condensate (fig.~\ref{2point-fig}) leaving behind only two fermion lines. The IR scaling dimension of any such operator will thus be the same and independent of $L$. Though this is  a tree-level argument, it is plausible that the degeneracy remains to all orders in perturbation theory, and the IR dimension of any operator of this type reduces to that of the basic fermion bilinear.

\section{Conclusions}

At large distances the strongly coupled $\mathcal{N}=2^*$ SYM behaves as a five-dimensional CFT \cite{HoyosBadajoz:2010td}. 
We have argued that this unexpected result is a non-perturbative manifestation of Eguchi-Kawai mechanism. 
The key point is the flattening of the eigenvalue density of the symmetry-breaking scalar VEV. The color average then mimics integration of the fifth component of momentum. Considering the quark-anti-quark potential as an example, we were able to reproduce the right dependence on the 't~Hooft coupling and the distance between quarks, $V(\lambda )\sim \sqrt{\lambda }/L^2$, by just plugging the strong-coupling eigenvalue density in the tree-level potential. The constant of proportionality comes out wrong, which is not surprising as our calculation is not really self-consistent. The eigenvalue density is exactly known from the supergravity analysis \cite{Buchel:2000cn} and supersymmetric localization \cite{Buchel:2013id}, but keeping only tree level diagrams is certainly not a good approximation at strong coupling. It would be interesting to see if the tree-level truncation can be improved, for example by resumming infinite classes of diagrams.

Another unusual feature of the holographic dual of $\mathcal{N}=2^*$ theory is a huge degeneracy of IR scaling dimensions. A qualitative explanation for this fact is that scalar lines in Feynman diagrams can be absorbed into the Higgs condensate, making operators that differ by their scalar content degenerate in the IR. The argument is again perturbative, and its validity at strong coupling requires an independent confirmation. It is desirable in this respect to promote our rather sketchy calculation of scaling dimensions on the holographic side to a full-fledged analysis of linear perturbations in the PW background.

\subsection*{Acknowledgments}

We would like to thank E.~Kiritsis for discussions.
The work of K.Z. was supported by the ERC advanced grant No 341222, by the Marie
Curie network GATIS of the European Union's FP7 Programme under REA Grant
Agreement No 317089, and by the Swedish Research Council (VR) grant
2013-4329. DY acknowledges NORDITA where this work was begun, during
his time as a NORDITA fellow.


\providecommand{\href}[2]{#2}\begingroup\raggedright\endgroup

\end{document}